\pdfoutput=1
\RequirePackage{ifpdf}
\ifpdf 
\documentclass[pdftex]{sigma}
\else
\documentclass{sigma}
\fi

{\theoremstyle{definition}

}

\begin{document}

\allowdisplaybreaks

\renewcommand{\PaperNumber}{035}

\FirstPageHeading

\ShortArticleName{On Addition Formulae of KP, mKP and BKP Hierarchies}

\ArticleName{On Addition Formulae of KP, mKP\\
and BKP Hierarchies}

\Author{Yoko SHIGYO}

\AuthorNameForHeading{Y.~Shigyo}

\Address{Department of Mathematics, Tsuda College, Kodaira, Tokyo, 187-8577, Japan}
\Email{\href{mailto:yoko.shigyo@gmail.com}{yoko.shigyo@gmail.com}}

\ArticleDates{Received December 12, 2012, in f\/inal form April 04, 2013; Published online April 23, 2013}

\Abstract{In this paper we study the addition formulae of the KP, the mKP and the BKP hierarchies.
We prove that the total hierarchies are equivalent to the simplest equations of their addition formulae.
In the case of the KP and the mKP hierarchies those results had previously been proved by Noumi, Takasaki and Takebe by
way of wave functions.
Here we give alternative and direct proofs for the case of the KP and mKP hierarchies.
Our method can equally be applied to the BKP hierarchy.}


\Classification{14H70; 37K10; 37K20}

\section{Introduction}

Our purpose is to prove that some integrable hierarchies are equivalent to the simplest
equations of their addition formulae.
In this paper  we study the KP, the modif\/ied KP (mKP) and the BKP hierarchies.

The (bilinear) KP hierarchy~\cite{JM1} is an inf\/inite system of bilinear equations for $\tau(x)$,
$x=(x_1,x_2,\dots)$, given, in the generating form, by
\begin{gather*}
\oint e^{-2\xi(y,\lambda)}\tau\big(x-y-\big[\lambda^{-1}\big]\big)\tau\big(x+y+\big[\lambda^{-1}\big]\big)\frac{d\lambda}{2\pi i}=0.
\end{gather*}
Namely, if we expand the left hand side in $y=(y_1,y_2,\dots)$, then we get Hirota's bilinear equations,
which contain, as the simplest equation, the Kadomtsev--Petviashvili (KP) equation in the bilinear form
\begin{gather*}
\big(D_1^4+3D_2^2-4D_1D_3\big)\tau\cdot\tau=0,
\end{gather*}
where $D_i$ is the Hirota's bilinear operator def\/ined by
\begin{gather*}
\big(D_1^{i_1}D_2^{i_2}\cdots\big)\tau\cdot\tau=\left(\left(\frac{\partial}{\partial y_1}\right)^{i_1}
\left(\frac{\partial}{\partial y_2}\right)^{i_2}\cdots\right)(\tau(x+y)\tau(x-y))\Bigl|_{y=0},
\qquad
y=(y_1,y_2,\dots).
\end{gather*}
If we put $y=\left(\sum\limits_{i=1}^{m-1}[\beta_i]-\sum\limits_{i=1}^{m+1}[\alpha_i]\right)/2$, $[\alpha]=(\alpha,\alpha^2/2,\alpha^3/3,\dots)$, instead of expanding in $y$,
and compute the integral by taking residues, then we get
addition formulae~\cite{SS1}:
\begin{gather}
\sum_{i=1}^{m+1}(-1)^{i-1}C_{\alpha\beta}\tau
\left(x+\sum_{j=1}^{m-1}[\beta_j]+[\alpha_i]\right)\tau
\left(x+\sum_{j=1,j\neq i}^{m+1}[\alpha_j]\right)=0,
\qquad  m\ge2,
\label{eq:204}
\end{gather}
where $C_{\alpha\beta}$ depend only on $\{\alpha_i\}_{1\le i\le m+1}$ and $\{\beta_i\}_{1\le i\le m-1}$.
The simplest case of addition formulae is the case of $m$=2:
\begin{gather}
\alpha_{12}\alpha_{34}\tau(x+[\alpha_1]+[\alpha_2])\tau(x+[\alpha_3]+[\alpha_4])
-\alpha_{13}\alpha_{24}\tau(x+[\alpha_1]+[\alpha_3])\tau(x+[\alpha_2]+[\alpha_4])
\nonumber
\\
\qquad
{}+\alpha_{14}\alpha_{23}\tau(x+[\alpha_1]+[\alpha_4])\tau(x+[\alpha_2]+[\alpha_3])=0,
\label{eq:200}
\end{gather}
where $\alpha_{ij}=\alpha_i-\alpha_j$.
It is surprising that the KP hierarchy itself is equivalent to~\eqref{eq:200}.
This fact has been proved by Takasaki and Takebe~\cite{TT1} by way of the wave functions of the KP
hierarchy.
Here we give an alternative and direct proof.
First we show that the totality of addition formulae~\eqref{eq:204} is equivalent to the KP hierarchy by
using the properties of symmetric functions.
If we call the function of the form $\tau(x+[\alpha_1]+\cdots+[\alpha_n])$ the $n$-point function,
then~\eqref{eq:200} is a~relation among two point functions.
By shifting $x$ appropriately we can consider~\eqref{eq:200} as an expression of a~four point function in
terms of two point functions (see Proposition~\ref{proposition2}).
Repeating this process, we can derive the formulae which express the 2$m$-point function as a~determinant
of 2-point functions.
These formulae are called Fay's determinant formulae in the case of theta function~\cite{F1,R1}.
In~\cite{F1} it is indicated without proofs that the determinant formulae can be obtained from the
trisecant formulae corresponding to~\eqref{eq:200}.
In this sense the determinant formulae~\eqref{eq:12} and their derivation from~\eqref{eq:200} cannot be
considered a~new result\footnote{Takasaki K., Private communications.}. 
Next we show that the Pl\"{u}cker's relations for the determinants appearing in these formulae are nothing
but the addition formulae~\eqref{eq:204} for $m$-point functions.
In this way, we can prove that~\eqref{eq:200} is equivalent to the KP hierarchy.
For the mKP and the BKP hierarchies, similar results hold although there are some dif\/ferences.

The mKP hierarchy~\cite{JM1} is an inf\/inite system of dif\/ferential equations for an inf\/inite number
of functions $\tau_l(x)$, $l\in\mathbb{Z}$.
In this case there are an inf\/inite number of the simplest addition formulae:
\begin{gather}
\alpha_{23}\tau_l(x+[\alpha_1])\tau_{l+1}(x+[\alpha_2]+[\alpha_3])
-\alpha_{13}\tau_l(x+[\alpha_2])\tau_{l+1}(x+[\alpha_1]+[\alpha_3])
\nonumber
\\
\qquad
{}+\alpha_{12}\tau_l(x+[\alpha_3])\tau_{l+1}(x+[\alpha_1]+[\alpha_2])=0,\qquad  l\in\mathbb{Z}.
\label{eq:201}
\end{gather}
It had been proved that~\eqref{eq:201} is equivalent to the mKP hierarchy in~\cite{NT1}.
Here we prove the equivalence in a~similar strategy to the case of KP.
A new feature of the present case is that there exist addition formulae involving $\tau_l$ and $\tau_{l+k}$
for $k\ge2$.
We prove that these addition formulae are consequences of~\eqref{eq:201}.

The BKP hierarchy is an inf\/inite system of bilinear equations for $\tau(x)$, $x=(x_1,x_3,\dots)$.
The following is the simplest addition formula which has four terms:
\begin{gather}
\tilde{\alpha}_{12}\tilde{\alpha}_{13}\alpha_{23}
\tau(x+2[\alpha_1]_{\rm o})\tau(x+2[\alpha_2]_{\rm o}+2[\alpha_3]_{\rm o})
\nonumber
\\
\qquad
{}-\tilde{\alpha}_{12}\tilde{\alpha}_{23}\alpha_{13}
\tau(x+2[\alpha_2]_{\rm o})\tau(x+2[\alpha_1]_{\rm o}+2[\alpha_3]_{\rm o})\nonumber
\\
\qquad
{}+\tilde{\alpha}_{13}\tilde{\alpha}_{23}\alpha_{12}
\tau(x+2[\alpha_3]_{\rm o})\tau(x+2[\alpha_1]_{\rm o}+2[\alpha_2]_{\rm o})\nonumber
\\
\qquad
{}-\alpha_{12}\alpha_{13}\alpha_{23}\tau(x)\tau(x+2[\alpha_1]_{\rm o}+2[\alpha_2]_{\rm o}+2[\alpha_3]_{\rm o})=0,
\label{eq:202}
\end{gather}
where $\tilde{\alpha}_{ij}=\alpha_i+\alpha_j$ and $[\alpha]_{\rm o}=\big(\alpha,\alpha^3/3,\alpha^5/5,\dots\big)$.
We prove that~\eqref{eq:202} is equivalent to the BKP hierarchy in a~similar way to the KP hierarchy.
In this case we use Pfaf\/f\/ians instead of determinants to express $n$-point functions in terms of one
and two point functions.
To this end we need the analogue of Sylvester's theorem and the Pl\"{u}cker's relations for
Pfaf\/f\/ians~\cite{H2,O1}.

We have shown that the KP, the mKP and the BKP hierarchies are equivalent to the simplest addition formulae.
It is interesting to study whether, for other integrable hierarchies~\cite{DJM1-I,DJM1-II,DJM1-V,JM1,T1,T2}, similar
structure exists.
There exists a~result for the Toda hierarchy~\cite{T1,Te1}.
But the problem arises to specify what are the fundamental equations in general.
To consider these problems, it will be ef\/fective to use free fermion descriptions of integrable
hierarchies~\cite{DJM1-I,DJM1-II,DJM1-V,JM1}.
It is also interesting to apply the results to the study of discrete dif\/ferential
geometries~\cite{BS1, IKMO1} and addition formulae for sigma functions~\cite{BEL1,EEG1,N1}.

This paper consists of three sections and three appendices.
In Section~\ref{section2}, we consider the KP hierarchy.
The key point is to~prove the equivalence between the KP hierarchy and its inf\/inite sequence of
addition formulae.
Since this case is fundamental, the details are given.
Then we consider the mKP hierarchy in Section~\ref{section3}.
The arguments which are similar to the KP hierarchy are omitted.
In Section~\ref{section4}, we study the BKP hierarchy.
The Pfaf\/f\/ians are necessary in this case.
Necessary properties of Pfaf\/f\/ians including the def\/inition are reviewed in Appendices~\ref{appendixA},~\ref{appendixB} and~\ref{appendixC}.

\section[The addition formula for the $\tau$-function of the KP hierarchy]{The addition formula for the $\boldsymbol{\tau}$-function of the KP hierarchy} \label{section2}

Let
\begin{gather*}
[\alpha]=\left(\alpha,\frac{\alpha^2}{2},\frac{\alpha^3}{3},\dots\right),
\qquad
\xi(t,\lambda)=\sum_{n=1}^{\infty}t_n\lambda^n,
\qquad
t=(t_1,t_2,t_3,\dots).
\end{gather*}
The KP hierarchy is a~system of equations for a~function $\tau(t)$ \cite{DJKM1,JM1, MJD1} given by
\begin{gather}
\oint e^{\xi(t'-t,\lambda)}\tau\big(t'-\big[\lambda^{-1}\big]\big)\tau\big(t+\big[\lambda^{-1}\big]\big)\frac{d\lambda}{2\pi i}=0.
\label{eq:1}
\end{gather}
Here $\oint$ means a~formal algebraic operator extracting the coef\/f\/icient of $z^{-1}$ of Laurent series:
\begin{gather*}
\oint\frac{dz}{2\pi i}\sum_{n=-\infty}^{\infty}a_n z^n=a_{-1}.
\end{gather*}
Set $t=x+y$, $t'=x-y$.
Then~\eqref{eq:1} becomes
\begin{gather}
\oint e^{-2\xi(y,\lambda)}\tau\big(x-y-\big[\lambda^{-1}\big]\big)\tau\big(x+y+\big[\lambda^{-1}\big]\big)\frac{d\lambda}{2\pi i}=0.
\label{eq:3}
\end{gather}
For an integer $m\ge2$, set
\begin{gather}
y=\frac{1}{2}\left(\sum_{i=1}^{m-1}[\beta_i]-\sum_{i=1}^{m+1}[\alpha_i]\right)
\label{eq:22}
\end{gather}
in~\eqref{eq:3}. Then it becomes
\begin{gather}
\oint\exp\left({-\xi\left(\displaystyle\sum_{i=1}^{m-1}[\beta_i]-\sum_{i=1}^{m+1}[\alpha_i],\lambda\right)}\right)
\tau\left(x-\frac{1}{2}\left(\sum_{i=1}^{m-1}[\beta_i]-\sum_{i=1}^{m+1}[\alpha_i]\right)-\big[\lambda^{-1}
\big]\right)\nonumber
\\
\qquad
{}\times\tau\left(x+\frac{1}{2}\left(\sum_{i=1}^{m-1}[\beta_i]-\sum_{i=1}^{m+1}[\alpha_i]\right)+\big[\lambda^{-1}
\big]\right)\frac{d\lambda}{2\pi i}=0.
\label{eq:4}
\end{gather}
By virtue of the identity
\begin{gather*}
\sum_{n=1}^{\infty}\frac{x^n}{n}=-\log(1-x),
\end{gather*}
the exponential factor in~\eqref{eq:4} reduces to a~rational function of $\lambda$, $\alpha_i$, $\beta_i$ as
\begin{gather*}
\exp\left(-\xi\left(\sum_{i=1}^{m-1}[\beta_i]-\sum_{i=1}^{m+1}[\alpha_i],\lambda\right)\right)
=\frac{\prod\limits_{i=1}^{m-1}(1-\beta_i\lambda)}{\prod\limits_{i=1}^{m+1}
(1-\alpha_i\lambda)}.
\end{gather*}
Computing the integral by taking residues at $\lambda=\alpha_i^{-1}$, $1\le i\le m+1$ \cite{M1} and
shifting the variable $x$ as
\begin{gather*}
x\rightarrow x+\frac{1}{2}\left(\sum_{i=1}^{m-1}[\beta_i]+\sum_{i=1}^{m+1}[\alpha_i]\right),
\end{gather*}
we get the following addition formulae of the $\tau$-function~\cite{SS1}
\begin{gather}
\sum_{i=1}^{m+1}(-1)^{i-1}\zeta(x;\beta_1,\dots,\beta_{m-1},\alpha_i)\zeta(x;\alpha_1,\dots,\hat{\alpha}
_i,\dots,\alpha_{m+1})=0,\qquad  m\ge2,
\label{eq:5}
\end{gather}
where
\begin{gather*}
\zeta(x;\alpha_1,\dots,\alpha_{n})=\Delta(\alpha_1,\dots,\alpha_{n})\tau(x+[\alpha_1]+\dots+[\alpha_{n}]),
\\
\Delta(\alpha_1,\dots,\alpha_n)=\prod_{i<j}(\alpha_i-\alpha_j),
\end{gather*}
and $\hat{\alpha}_i$ means that $\alpha_i$ should be removed.
\begin{example}\label{example1}
In the case of $m=2$, we have
\begin{gather}
\alpha_{12}\alpha_{34}\tau(x+[\alpha_1]+[\alpha_2])\tau(x+[\alpha_3]+[\alpha_4])
-\alpha_{13}\alpha_{24}\tau(x+[\alpha_1]+[\alpha_3])\tau(x+[\alpha_2]+[\alpha_4])
\nonumber
\\
\qquad
{}+\alpha_{14}\alpha_{23}\tau(x+[\alpha_1]+[\alpha_4])\tau(x+[\alpha_2]+[\alpha_3])=0,
\label{eq:6}
\end{gather}
where $\alpha_{ij}=\alpha_i-\alpha_j$.
\end{example}

We call~\eqref{eq:6} `the three term equation'.
We have derived~\eqref{eq:6} from~\eqref{eq:1}.
The fact that the converse is true is proved by Takasaki and Takebe~\cite{TT1}.
\begin{theorem}[\cite{TT1}] \label{theorem1} The three term equation~\eqref{eq:6} is equivalent to the KP hierarchy~\eqref{eq:1}.
\end{theorem}

In~\cite{TT1} the theorem is proved by constructing the wave function of the KP-hierarchy.
To do it the dif\/ferential Fay identity, which is a~certain limit of~\eqref{eq:6}, is used.
Here we give an alternative and direct proof of the theorem.
\begin{proposition}\label{proposition1}
The KP hierarchy~\eqref{eq:1} is equivalent to~\eqref{eq:5}.
\end{proposition}
\begin{proof}
It is suf\/f\/icient to prove that if~\eqref{eq:4} holds for any $m\ge2$ and arbitrary
$\alpha_i\neq0$, $\beta_i$, then~\eqref{eq:3} holds.
Set the left hand side of~\eqref{eq:3} to be $F(y)$.
Expand $F(y)$ in $y$ as
\begin{gather*}
F(y)=\sum_\gamma F_\gamma y^\gamma,\qquad  y^\gamma=y_1^{\gamma_1}y_2^{\gamma_2}
\cdots,\qquad \gamma=(\gamma_1,\gamma_2,\dots).
\end{gather*}
We consider the case $\beta_i=0$, $1\le i\le m-1$ in~\eqref{eq:22} and set
\begin{gather*}
y'=\sum_{i=1}^{m+1}[\alpha_i].
\end{gather*}
We prove $F_\gamma=0$ for any $\gamma$ if $F\big({-}\frac{y'}{2}\big)=0$ for any $m\ge2$.
We consider $m$ f\/ixed.
Let us def\/ine the weight of $y_i$ to be $i$ and ${\rm wt}\,y^\gamma=\sum\limits_{i=1}^{\infty}i\gamma_i$.
Decompose $F$ according to weights as
\begin{gather*}
F=F^{(0)}+F^{(1)}+F^{(2)}+\cdots,
\qquad
F^{(i)}=\sum_{{\rm wt}\,y^\gamma=i}F_\gamma y^\gamma.
\end{gather*}
We substitute $-\frac{y'}{2}$ to $y$ and get the homogeneous polynomial of degree $i$ of
$\alpha_1,\dots,\alpha_{m+1}$:
\begin{gather*}
F^{(i)}\left(-\frac{y'}{2}\right)=\sum_{\gamma_1+\dots+\gamma_{m+1}=i}b^{(i)}_{\gamma_1\dots\gamma_{m+1}}
\alpha_1^{\gamma_1}\cdots\alpha_{m+1}^{\gamma_{m+1}}.
\end{gather*}
Then $F\big({-}\frac{y'}{2}\big)=0$ is equivalent to $F^{(i)}\big({-}\frac{y'}{2}\big)=0$ for any $i$.
Notice that $i y'_i=\alpha_1^i+\cdots+\alpha_{m+1}^i$ is a~power sum symmetric function.
Therefore $y_1',\dots,y_{m+1}'$ are algebraically independent (see~\cite[(2.12)]{Mac}).
If $i\le m+1$, then $F^{(i)}(y)$ is a~polynomial at most of $y_1,\dots,y_{m+1}$.
Thus $F^{(i)}\big({-}\frac{y'}{2}\big)=0$ implies $F_\gamma=0$ for any $\gamma$ satisfying ${\rm wt}\,\gamma=i$.
Since $m$ is arbitrary we have $F_\gamma=0$ for any $\gamma$.
\end{proof}

\begin{remark}\label{remark1}
In the course of the proof we actually prove the equivalence between~\eqref{eq:1} and~\eqref{eq:5} with
$\beta_i=0$ for any $i$ (of course, in that case we have f\/irstly to divide~\eqref{eq:5} by
$\Delta(\beta_1,\dots,\beta_{m-1})$).
\end{remark}
\begin{proposition}\label{proposition2}
The following formula follows from~\eqref{eq:6}:
\begin{gather}
\frac{\tau\left(x+\sum\limits_{i=1}^{m}[\beta_i]-\sum\limits_{i=1}^{m}[\alpha_i]\right)}{\tau(x)}
\nonumber
\\
\qquad
=\frac{\prod\limits_{i,j=1}^{m}
(\beta_i-\alpha_j)}{\prod\limits_{i<j}\alpha_{ij}\beta_{ji}}\det\left(\frac{\tau(x+[\beta_i]-[\alpha_j])}
{(\beta_i-\alpha_j)\tau(x)}\right)_{1\le i,j\le m},
\qquad
m\ge2.
\label{eq:12}
\end{gather}
\end{proposition}
\begin{proof}
We change the names of $\alpha$ variables in~\eqref{eq:6} as
$(\alpha_3,\alpha_4)\rightarrow(\beta_1,\beta_2)$ and shift $x$ to $x\rightarrow x-[\alpha_1]-[\alpha_2]$.
After that we solve it in $\tau(x+[\beta_1]+[\beta_2]-[\alpha_1]-[\alpha_2])$ we get $m=2$ case
of~\eqref{eq:12}.

Suppose that~\eqref{eq:12} holds in case of $m=k$:
\begin{gather}
\tau\left(x+\sum_{i=1}^{k}[\beta_i]-\sum_{i=1}^{k}[\alpha_i]\right)
=\tau(x)^{-k+1}C_k\det\left(\frac{\tau(x+[\beta_i]-[\alpha_j])}{\beta_i-\alpha_j}\right)_{1\le i,j\le k},
\label{eq:14}
\end{gather}
where
\begin{gather*}
C_k=\frac{\prod\limits_{i,j=1}^{k}(\beta_i-\alpha_j)}{\prod\limits_{i<j}\alpha_{ij}\beta_{ji}}.
\end{gather*}
In~\eqref{eq:14} we shift the variable $x$ as
\begin{gather*}
x\rightarrow x+[\beta_{k+1}]-[\alpha_{k+1}].
\end{gather*}
Then
\begin{gather}
\tau\left(x+\sum_{i=1}^{k+1}[\beta_i]-\sum_{i=1}^{k+1}[\alpha_i]\right)
=\tau(x+[\beta_{k+1}]-[\alpha_{k+1}])^{-k+1}C_k
\nonumber\\
\qquad
{}\times
\det\left(\frac{\tau(x+[\beta_i]+[\beta_{k+1}
]-[\alpha_j]-[\alpha_{k+1}])}{\beta_i-\alpha_j}\right)_{1\le i,j\le k}.
\label{eq:15}
\end{gather}
By~\eqref{eq:12} with $m=2$,
\begin{gather}
\tau(x+[\beta_i]+[\beta_{k+1}]-[\alpha_j]-[\alpha_{k+1}])=\tau(x)^{-1}A_{ij}\cdot X_{ij},
\label{eq:16}
\end{gather}
where
\begin{gather*}
A_{ij}=\frac{(\beta_i-\alpha_j)(\beta_i-\alpha_{k+1})(\beta_{k+1}-\alpha_j)(\beta_{k+1}-\alpha_{k+1})}
{(\alpha_j-\alpha_{k+1})(\beta_{k+1}-\beta_i)},
\\
X_{ij}=\det\left(
\begin{matrix}\dfrac{\tau(x+[\beta_i]-[\alpha_j])}{\beta_i-\alpha_j}&\dfrac{\tau(x+[\beta_i]-[\alpha_{k+1}])}
{\beta_i-\alpha_{k+1}}
\vspace{2mm}\\
\dfrac{\tau(x+[\beta_{k+1}]-[\alpha_j])}{\beta_{k+1}-\alpha_j}&\dfrac{\tau(x+[\beta_{k+1}]-[\alpha_{k+1}])}
{\beta_{k+1}-\alpha_{k+1}}
\end{matrix}
\right).
\end{gather*}
By substituting~\eqref{eq:16} to the determinant in the right hand side of~\eqref{eq:15}, we have
\begin{gather}
\det\left(\frac{\tau(x+[\beta_i]+[\beta_{k+1}]-[\alpha_j]-[\alpha_{k+1}])}{\beta_i-\alpha_j}
\right)_{1\le i,j\le k}\nonumber
\\
\qquad
{}=\left(\frac{\beta_{k+1}-\alpha_{k+1}}{\tau(x)}\right)^k\prod_{i=1}^{k}\frac{(\beta_{k+1}
-\alpha_i)(\beta_i-\alpha_{k+1})}{(\alpha_i-\alpha_{k+1})(\beta_{k+1}-\beta_i)}\det(X_{ij})_{1\le i,j\le k}.
\label{eq:17}
\end{gather}
Using Sylvester's theorem (Appendix~\ref{appendixB}),
\begin{gather}
\det(X_{ij})_{1\le i,j\le k}=\left(\frac{\tau(x+[\beta_{k+1}]-[\alpha_{k+1}])}{\beta_{k+1}-\alpha_{k+1}}
\right)^{k-1}\det\left(\frac{\tau(x+[\beta_i]-[\alpha_j])}{\beta_i-\alpha_j}\right)_{1\le i,j\le k+1}.
\label{eq:18}
\end{gather}
Substituting~\eqref{eq:18} and~\eqref{eq:17} into~\eqref{eq:15}, we get the case of $m=k+1$
of~\eqref{eq:12}.
\end{proof}

Let us consider an $N\times m$ matrix ${\mathcal A} =(a_{ij})_{1\le i\le N,1\le j\le m}$ with $N\ge m$ and
set, for $1\le l_1,\dots,l_m\le N$,
\begin{gather*}
A(l_1,\dots,l_m)=\det(a_{{l_i},j})_{1\le i,j\le m}.
\end{gather*}
For any $1\le k_1,\dots,k_{m-1},l_1,\dots,l_{m+1}\le N$ these determinants satisfy Pl\"{u}cker's
relations:
\begin{gather}
\sum_{i=1}^{m+1}(-1)^{i-1}A(k_1,\dots,k_{m-1},l_i)A(l_1,\dots,{\hat l}_i,\dots,l_{m+1})=0.
\label{eq:20}
\end{gather}
\begin{proposition}\label{proposition3}
The Pl\"{u}cker's relations for the determinant of the right hand side of~\eqref{eq:12} give the addition
formulae~\eqref{eq:5}.
\end{proposition}
\begin{proof}
Let $m$ be f\/ixed.
Consider the $\infty\times m$ matrix ${\mathcal A}=(a_{ij})$ with
\begin{gather*}
a_{ij}=\frac{\tau(x+[\beta_i]-[\alpha_j])}{(\beta_i-\alpha_j)\tau(x)}.
\end{gather*}
Then $A(k_1,\dots,k_{m-1},l_i)$ and $A(l_1,\dots,{\hat l}_i,\dots,l_{m+1})$ can be expressed by $2m$ point
functions by~\eqref{eq:12}.
We substitute them to~\eqref{eq:20} and shift the variable $x$ as
\begin{gather*}
x\rightarrow x+\sum_{s=1}^m[\alpha_s].
\end{gather*}
Then we get the additional 
formulae~\eqref{eq:5} by renaming the variables as
$(\beta_{k_1},\dots,\beta_{k_{m-1}})\rightarrow(\beta_1,\dots,\beta_{m-1})$,
$(\beta_{l_1},\dots,\beta_{l_{m+1}})\rightarrow(\alpha_1,\dots,\alpha_{m+1})$.
\end{proof}

\begin{proof}[Proof of Theorem \ref{theorem1}.] By Propositions \ref{proposition1}, \ref{proposition2} and \ref{proposition3}, we have~\eqref{eq:1} from~\eqref{eq:6}.
Thus Theorem~\ref{theorem1} is proved.
\end{proof}

\section{mKP hierarchy}\label{section3}

Let $\tau_l(t)$ ($l\in\mathbb{Z}$) be $\tau$-functions of the modif\/ied KP (mKP) hierarchy.
We use the same notation as that for KP hierarchy ($[\alpha]$, $\xi(t,\lambda)$, etc.).

The mKP hierarchy is given by the bilinear equation \cite{DJKM1,JM1} of the form
\begin{gather}
\oint e^{\xi(t'-t,\lambda)}\lambda^{l-l'}\tau_l\big(t'-\big[\lambda^{-1}\big]\big)\tau_{l'}
\big(t+\big[\lambda^{-1}\big]\big)\frac{d\lambda}
{2\pi i}=0,\qquad l\ge l'.
\label{eq:30}
\end{gather}
Set $t=x+y$, $t'=x-y$.
Then~\eqref{eq:30} becomes
\begin{gather}
\oint e^{-2\xi(y,\lambda)}\lambda^{l-l'}\tau_l\big(x-y-\big[\lambda^{-1}\big]\big)
\tau_{l'}\big(x+y+\big[\lambda^{-1}\big]\big)\frac{d\lambda}{2\pi i}=0,\qquad l\ge l'.
\label{eq:31}
\end{gather}
Let $l-l'=k\ge0$.
Set
\begin{gather*}
y=\frac{1}{2}\left(\sum_{i=1}^{m-2}[\beta_i]-\sum_{i=1}^{m+k}[\alpha_i]\right).
\end{gather*}
Then the exponential factor in~\eqref{eq:31} reduces to a~rational function of $\lambda$, $\alpha_i$, $\beta_i$
as in the KP case:
\begin{gather*}
\exp\left(-\xi\left(\sum_{i=1}^{m-2}[\beta_i]-\sum_{i=1}^{m+k}
[\alpha_i],\lambda\right)\right)=\frac{\prod\limits_{i=1}^{m-2}(1-\beta_i\lambda)}{\prod\limits_{i=1}^{m+k}
(1-\alpha_i\lambda)}.
\end{gather*}
Computing the integral by taking residues at $\lambda=\alpha_i^{-1}$, $1\le i\le m+k$ and shifting the
va\-riab\-le~$x$ as
\begin{gather*}
x\rightarrow x+\frac{1}{2}\left(\sum_{i=1}^{m-2}[\beta_i]+\sum_{i=1}^{m+k}[\alpha_i]\right),
\end{gather*}
we have the following addition formulae of the mKP hierarchy:
\begin{gather}
\sum_{i=1}^{m+k}(-1)^{i-1}\zeta_l(x;\beta_1,\dots,\beta_{m-2},\alpha_i)\zeta_{l+k}
(x;\alpha_1,\dots,\hat{\alpha}_i,\dots,\alpha_{m+k})=0,
\nonumber
\\
\qquad
l\in\mathbb{Z},\qquad k\ge0,\qquad m\ge2,
\label{eq:32}
\end{gather}
where
\begin{gather*}
\zeta_l(x;\alpha_1,\dots,\alpha_n)=\Delta(\alpha_1,\dots,\alpha_n)\tau_l
\left(x+\sum_{i=1}^n[\alpha_i]\right).
\end{gather*}
\begin{example}\label{example2}
The case $l-l'=1$ and $m=2$ of~\eqref{eq:32} is
\begin{gather}
\alpha_{23}\tau_l(x+[\alpha_1])\tau_{l+1}(x+[\alpha_2]+[\alpha_3])
-\alpha_{13}\tau_l(x+[\alpha_2])\tau_{l+1}(x+[\alpha_1]+[\alpha_3])\nonumber
\\
\qquad
{}+\alpha_{12}\tau_l(x+[\alpha_3])\tau_{l+1}(x+[\alpha_1]+[\alpha_2])=0.
\label{eq:33}
\end{gather}
\end{example}

We call~\eqref{eq:33} `the three term equation of the mKP hierarchy'.

The following theorem is proved in~\cite{NT1}.
\begin{theorem}[\cite{NT1}]\label{theorem2} The three term equation~\eqref{eq:33} is equivalent to the mKP hierarchy~\eqref{eq:30}.
\end{theorem}

We give an another proof which is similar to that of the KP hierarchy.
The following proposition can be proved as in the KP-case.
\begin{proposition}\label{proposition4}
The mKP hierarchy~\eqref{eq:30} is equivalent to~\eqref{eq:32}.
\end{proposition}
\begin{proposition}\label{proposition5}
The following formula follows from~\eqref{eq:33} for $n\ge2$:
\begin{gather}
\frac{\tau_{l+1}\left(x+\sum\limits_{i=1}^{n}[\alpha_i]-\sum\limits_{i=1}^{n-1}[\beta_i]\right)}{\tau_l(x)}\nonumber
\\
\qquad
{}=C\det\left(
\begin{matrix}\dfrac{\tau_l(x+[\alpha_1]-[\beta_1])}{(\alpha_1-\beta_1)\tau_l(x)}
&\cdots&\dfrac{\tau_l(x+[\alpha_1]-[\beta_{n-1}])}{(\alpha_1-\beta_{n-1})\tau_l(x)}&\dfrac{\tau_{l+1}
(x+[\alpha_1])}{\tau_l(x)}
\vspace{2mm}\\
\dfrac{\tau_l(x+[\alpha_2]-[\beta_1])}{(\alpha_2-\beta_1)\tau_l(x)}
&\cdots&\dfrac{\tau_l(x+[\alpha_2]-[\beta_{n-1}])}{(\alpha_2-\beta_{n-1})\tau_l(x)}&\dfrac{\tau_{l+1}
(x+[\alpha_2])}{\tau_l(x)}
\\
\vdots&\ddots&\vdots&\vdots
\\
\dfrac{\tau_l(x+[\alpha_n]-[\beta_1])}{(\alpha_n-\beta_1)\tau_l(x)}
&\cdots&\dfrac{\tau_l(x+[\alpha_n]-[\beta_{n-1}])}{(\alpha_n-\beta_{n-1})\tau_l(x)}&\dfrac{\tau_{l+1}
(x+[\alpha_n])}{\tau_l(x)}
\end{matrix}
\right),
\label{eq:34}
\end{gather}
where
\begin{gather*}
C=C(l,\{\alpha_i\},\{\beta_i\})=\frac{\prod\limits_{i=1}^{n}\prod\limits_{j=1}^{n-1}(\alpha_i-\beta_j)}{\left(\prod\limits_{i<j}^{n-1}
\beta_{ij}\right)\left(\prod\limits_{i>j}^{n}\alpha_{ij}\right)}.
\end{gather*}
\end{proposition}
\begin{proof}
The proof is similar to that of Proposition~\ref{proposition2}.
Therefore we leave details to readers.
\end{proof}
\begin{proposition}\label{proposition6}
The Pl\"{u}cker's relations for the determinant of the right hand side of~\eqref{eq:34} give~\eqref{eq:32}
with $k=1$.
\end{proposition}

\begin{proof}
For $m\ge1$ consider the $m\times2m$ matrix ${\mathcal A}=(a_{ij})_{1\le i\le m,\,1\le j\le2m}$ given by
\begin{gather*}
a_{ij}=\frac{\tau_l(x+[\alpha_i]-[\beta_j])}{\alpha_i-\beta_j},\qquad 1\le j\le2m-1,
\qquad
a_{i,2m}=\tau_{l+1}(x+[\alpha_i]).
\end{gather*}
For $1\le r_1,\dots,r_m\le2m$ we set
\begin{gather*}
A(r_1,\dots,r_m)=\det(a_{i,{r_j}})_{1\le i,j\le m}.
\end{gather*}
Then the Pl\"{u}cker's relation gives $k=1$ case of~\eqref{eq:32} by Propositions~\ref{proposition2} and~\ref{proposition5}.
\end{proof}

 {\sloppy By Propositions~\ref{proposition5} and~\ref{proposition6}, equation~\eqref{eq:32} with $k=1$ and arbitrary $m\ge2$ follows
from~\eqref{eq:33}.
The next lemma shows that~\eqref{eq:32} with $k\ge2$ and $m\ge2$ also follows from~\eqref{eq:33}.
The fact that~\eqref{eq:32} with $k=0$ follows from~\eqref{eq:33} is proved in~\cite{NT1}.
We generalize the arguments in~\cite{NT1} for $k\ge2$.

}

\begin{lemma}\label{lemma1}
Equation~\eqref{eq:32} follows from~\eqref{eq:33}.
\end{lemma}
\begin{proof}
We prove the lemma by induction on $k$.
Suppose that equation~\eqref{eq:32} is valid for $k$ and any $m\ge2$.
Shift the variable $x$ as
\begin{gather*}
x\rightarrow x-\sum_{j=1}^{m+k-1}[\alpha_j],
\end{gather*}
and multiply the resulting equation by $\tau_{l+k+1}(x+[\alpha_{m+k+1}])$.
Then we get
\begin{gather}
\sum_{i=1}^{m+k-1}A_i\tau_l\left(x-\sum_{j\neq i}^{m+k-1}[\alpha_j]+\sum_{j=1}^{m-2}[\beta_j]\right)\tau_{l+k}
\left(x+[\alpha_{m+k}]-[\alpha_i]\right)\tau_{l+k+1}(x+[\alpha_{m+k+1}])\nonumber
\\
\qquad
{}+A_{m+k}\tau_l\left(x+[\alpha_{m+k}]+\sum_{j=1}^{m-2}[\beta_j]-\sum_{j=1}^{m+k-1}[\alpha_j]\right)\tau_{l+k}
(x)\tau_{l+k+1}(x+[\alpha_{m+k}])=0,
\label{eq:36}
\end{gather}
where
\begin{gather*}
A_i=(-1)^{i-1}\Delta(\beta_1,\dots,\beta_{m-2},\alpha_i)\Delta(\alpha_1,\dots,\hat{\alpha}
_i,\dots,\alpha_{m+k}).
\end{gather*}
In~\eqref{eq:33} with $l$ being replaced by $l+k$, make a~shift $x\rightarrow x-[\alpha_3]$ and change the
label of $\alpha$ as $(\alpha_1,\alpha_2,\alpha_3)\rightarrow(\alpha_{m+k},\alpha_{m+k+1},\alpha_i)$, $1\le
i\le m+k-1$.
Then we get
\begin{gather}
\tau_{l+k}(x+[\alpha_{m+k}]-[\alpha_i])\tau_{l+k+1}(x+[\alpha_{m+k+1}])\nonumber
\\
\qquad
{}=\frac{\alpha_{m+k,i}}{\alpha_{m+k+1,i}}\tau_{l+k}(x+[\alpha_{m+k+1}]-[\alpha_i])\tau_{l+k+1}
(x+[\alpha_{m+k}])\nonumber
\\
\qquad\quad{}
-\frac{\alpha_{m+k,m+k+1}}{\alpha_{m+k+1,i}}\tau_{l+k}(x)\tau_{l+k+1}(x+[\alpha_{m+k}]+[\alpha_{m+k+1}
]-[\alpha_i]).
\label{eq:37}
\end{gather}
Substituting~\eqref{eq:37} to the summands of~\eqref{eq:36} and shifting $x$ as $x\rightarrow
x+\sum\limits_{j=1}^{m+k-1}[\alpha_j]$, then we get
\begin{gather}
\tau_{l+k+1}\left(x+\sum_{j=1}^{m+k}[\alpha_j]\right)
\sum_{i=1}^{m+k-1}A_i\frac{\alpha_{m+k,i}}{\alpha_{m+k+1,i}}
\tau_l\left(x+[\alpha_i]+\sum_{j=1}^{m-2}[\beta_j]\right)
\nonumber
\\
\qquad
{}\times\tau_{l+k}
\left(x+ \sum_{j\neq i}^{m+k-1}[\alpha_j]+[\alpha_{m+k+1}]\right)
+\tau_{l+k}\left(x+ \sum_{j=1}^{m+k-1}[\alpha_j]\right)
\nonumber
\\
\qquad
{}\times
\left\{\sum_{i=1}^{m+k-1}A_i\frac{\alpha_{m+k,m+k+1}}
{\alpha_{i,m+k+1}}\tau_l\left(x+[\alpha_i]+\sum_{j=1}^{m-2}[\beta_j]\right)\tau_{l+k+1}
\left(x+\sum_{j\neq i}^{m+k+1}[\alpha_j]\right)\right.
\nonumber
\\
\left.\qquad
{}+A_{m+k}\tau_l\left(x+[\alpha_{m+k}]+\sum_{j=1}^{m-2}[\beta_j]\right)
\tau_{l+k+1}\left(x+\sum_{j\neq m+k}^{m+k+1}[\alpha_j]\right)\right\}=0.
\label{eq:38}
\end{gather}
We write~\eqref{eq:32} in the form
\begin{gather}
\sum_{i=1}^{m+k-1}A_i\tau_l\left(x+[\alpha_i]+\sum_{j=1}^{m-2}[\beta_j]\right)\tau_{l+k}
\left(x+\sum_{j\neq i}^{m+k}[\alpha_j]\right)\nonumber
\\
\qquad
{}=-A_{m+k}\tau_l\left(x+[\alpha_{m+k}]+\sum_{j=1}^{m-2}[\beta_j]\right)\tau_{l+k}
\left(x+\sum_{j=1}^{m+k-1}[\alpha_j]\right).
\label{eq:39}
\end{gather}
Change $\alpha_{m+k}$ to $\alpha_{m+k+1}$ in~\eqref{eq:39}.
Notice that $A_i$, $i<m+k$, changes to
\begin{gather*}
(-1)^{i-1}\Delta(\beta_1,\dots,\beta_{m-2},\alpha_i)\Delta(\alpha_1,\dots,\hat{\alpha}
_i,\dots,\alpha_{m+k-1},\alpha_{m+k+1})
\\
\qquad
=\prod_{j=1}^{m+k-1}\frac{\alpha_{j,m+k+1}}{\alpha_{j,m+k}}
\cdot A_i\cdot\frac{\alpha_{i,{m+k}}}{\alpha_{i,{m+k+1}}}
\end{gather*}
and $A_{m+k}$ changes to
\begin{gather*}
(-1)^{m+k}\Delta(\beta_1,\dots,\beta_{m-2},\alpha_{m+k+1})\Delta(\alpha_1,\dots,\alpha_{m+k-1}).
\end{gather*}
Then we can rewrite the f\/irst term of~\eqref{eq:38} as
\begin{gather}
\tau_{l+k+1}\left(x+\sum_{j=1}^{m+k}[\alpha_j]\right)\sum_{i=1}^{m+k-1}A_i\frac{\alpha_{i,m+k}}{\alpha_{i,m+k+1}}
\tau_l\left(x+[\alpha_i]+\sum_{j=1}^{m-2}[\beta_j]\right)
\nonumber
\\
\qquad
{}\times
\tau_{l+k}\left(x+ \sum_{j\neq i}^{m+k-1}
[\alpha_j]+[\alpha_{m+k+1}]\right)
=\tau_{l+k+1}\left(x+\sum_{j=1}^{m+k}[\alpha_j]\right)B_{m+k+1}
\nonumber
\\
\qquad
{}\times\tau_l \left(x+[\alpha_{m+k+1}]+\sum_{j=1}^{m-2}
[\beta_j]\right)\tau_{l+k}\left(x+\sum_{j=1}^{m+k-1}[\alpha_j]\right),
\label{eq:40}
\end{gather}
where
\begin{gather*}
B_{m+k+1}=(-1)^{m+k+1}\frac{\Delta(\alpha_{m+k+1},\beta_1,\dots,\beta_{m-2}
)\Delta(\alpha_1,\dots,\alpha_{m+k})}{\prod\limits_{j=1}^{m+k-1}\alpha_{j,m+k+1}}.
\end{gather*}
Substitute~\eqref{eq:40} to~\eqref{eq:38} and divide the resulting equation by
$\tau_{l+k}\left(x+\sum\limits_{j=1}^{m+k-1}[\alpha_j]\right)$.
We, then, multiply it by $\prod\limits_{j=1}^{m+k-1}\alpha_{j,{m+k+1}}$ and get the case of $k+1$ of~\eqref{eq:32}.
\end{proof}

\section{BKP hierarchy}\label{section4}

Let $\tau(t)$ be the $\tau$-function of the BKP hierarchy.
In this case, the time variable is $t=(t_1,t_3,t_5,\dots)$.
We set
\begin{gather*}
[\alpha]_{\rm o}=\left(\alpha,\frac{\alpha^3}{3},\frac{\alpha^5}{5},\dots\right),\hskip1cm\tilde{\xi}
(t,\lambda)=\sum_{n=1}^{\infty}t_{2n-1}\lambda^{2n-1}.
\end{gather*}
The BKP hierarchy \cite{DJKM1,JM1} is def\/ined by
\begin{gather}
\oint e^{\tilde{\xi}(t-t',\lambda)}\tau\big(t-2\big[\lambda^{-1}\big]_{\rm o}\big)\tau\big(t'+2\big[\lambda^{-1}\big]_{\rm o}\big)\frac{d\lambda}
{2\pi i\lambda}=\tau(t)\tau(t').
\label{eq:50}
\end{gather}
Set $t=x-y$, $t'=x+y$.
We get
\begin{gather}
\oint e^{-2\tilde{\xi}(y,\lambda)}\tau\big(x-y-2\big[\lambda^{-1}\big]_{\rm o}\big)\tau\big(x+y+2\big[\lambda^{-1}\big]_{\rm o}\big)\frac{d\lambda}
{2\pi i\lambda}=\tau(x-y)\tau(x+y).
\label{eq:51}
\end{gather}
Set
\begin{gather*}
y=\sum_{i=1}^{n} [\alpha_i]_{\rm o}
\end{gather*}
in~\eqref{eq:51}. Then we have
\begin{gather*}
\oint e^{-2\tilde{\xi}\big(\sum\limits_{i=1}^{n} [\alpha_i]_{\rm o},\lambda\big)}\tau\left(x-\sum_{i=1}^{n}
 [\alpha_i]_{\rm o}-2\big[\lambda^{-1}\big]_{\rm o}\right)
\tau\left(x+\sum_{i=1}^{n} [\alpha_i]_{\rm o}+2\big[\lambda^{-1}\big]_{\rm o}\right)\frac{d\lambda}{2\pi i\lambda}
\\
\qquad
{}=\tau\left(x-\sum_{i=1}^{n} [\alpha_i]_{\rm o}\right)\tau\left(x+\sum_{i=1}^{n} [\alpha_i]_{\rm o}\right).
\end{gather*}
By decomposing $-2\sum\limits_{n=1}^\infty t_{2n-1}{\lambda}^{2n-1}$ as
\begin{gather*}
-2\sum_{n=1}^\infty t_{2n-1}{\lambda}^{2n-1}=-\sum_{n=1}^{\infty}t_n{\lambda}^n+\sum_{n=1}^{\infty}
t_n(-\lambda)^n,
\end{gather*}
we get
\begin{gather*}
\exp\left(-2\tilde{\xi}\left(\sum_{i=1}^{n} [\alpha_i]_{\rm o},\lambda\right)\right)=\prod_{i=1}^{n}
\frac{1-\alpha_i\lambda}{1+\alpha_i\lambda}.
\end{gather*}
Computing the integral by taking residues as before, shifting $x$ as $x+\sum\limits_{i=1}^{n}[\alpha_i]_{\rm o}$ and
divi\-ding by~$\tau(x)^2$ we have
\begin{gather}
\sum_{i=1}^{n}(-1)^{i-1}\frac{\tau(x+2[\alpha_i]_{\rm o})}{\tau(x)}A_{1\dots\hat{i}\dots n}^{-1}
\frac{\tau\left(x+2\sum\limits_{l=1,\, l\neq i}^n[\alpha_l]_{\rm o}\right)}{\tau(x)}\nonumber
\\
\qquad
{}-A_{1\dots n}^{-1}\frac{\tau\left(x+2\sum\limits_{l=1}^{n}[\alpha_l]_{\rm o}\right)}{\tau(x)}=0,\qquad  n \ \text{odd},
\label{eq:53}
\\
\sum_{i=1}^{n-1}(-1)^{i-1}\frac{\alpha_{i,n}}{\tilde{\alpha}_{i,n}}
\frac{\tau(x+2[\alpha_i]_{\rm o}+2[\alpha_n]_{\rm o})}{\tau(x)}A_{1\dots\hat{i}\dots n-1}^{-1}
\frac{\tau\left(x+2\sum\limits_{l=1,\, l\neq i}^{n-1}[\alpha_l]_{\rm o}\right)}{\tau(x)}\nonumber
\\
\qquad
{} -A_{1\dots n}^{-1}\frac{\tau\left(x+2\sum\limits_{l=1}^{n}[\alpha_l]_{\rm o}\right)}{\tau(x)}=0,\qquad  n \ \text{even}.
\label{eq:54}
\end{gather}
Here $A_{1\dots n}$ is def\/ined by
\begin{gather*}
A_{1\dots n}=\prod_{i<j}^{n}\frac{\tilde{\alpha}_{ij}}{\alpha_{ij}},\qquad \tilde{\alpha}_{ij}
=\alpha_i+\alpha_j,\qquad \alpha_{ij}=\alpha_i-\alpha_j.
\end{gather*}
\begin{example}\label{example3}
The case $n=3$ of~\eqref{eq:53} becomes
\begin{gather}
\frac{\tau\left(x+2\sum\limits_{i=1}^{3}[\alpha_i]_{\rm o}\right)}{\tau(x)}=A_{123}\left\{\frac{\tau(x+2[\alpha_1]_{\rm o})}{\tau(x)}
\frac{\alpha_{23}}{\tilde{\alpha}_{23}}\frac{\tau(x+2[\alpha_2]_{\rm o}+2[\alpha_3]_{\rm o})}{\tau(x)}\right.
\nonumber
\\
\hskip36mm-\frac{\tau(x+2[\alpha_2]_{\rm o})}{\tau(x)}\frac{\alpha_{13}}{\tilde{\alpha}_{13}}
\frac{\tau(x+2[\alpha_1]_{\rm o}+2[\alpha_3]_{\rm o})}{\tau(x)}\nonumber
\\
\hskip36mm\left.
+\frac{\tau(x+2[\alpha_3]_{\rm o})}{\tau(x)}\frac{\alpha_{12}}{\tilde{\alpha}_{12}}
\frac{\tau(x+2[\alpha_1]_{\rm o}+2[\alpha_2]_{\rm o})}{\tau(x)}\right\}.
\label{eq:55}
\end{gather}
\end{example}
We call equation~\eqref{eq:55} `the four term equation of the BKP hierarchy'.
\begin{example}
The case of $n=4$ of~\eqref{eq:54} is
\begin{gather}
\frac{\tau\left(x+2\sum\limits_{i=1}^{4}[\alpha_i]_{\rm o}\right)}{\tau(x)}
=A_{1234}\left\{\frac{\alpha_{14}}{\tilde{\alpha}_{14}}
\frac{\tau(x+2[\alpha_1]_{\rm o}+2[\alpha_4]_{\rm o})}{\tau(x)}\frac{\alpha_{23}}{\tilde{\alpha}_{23}}
\frac{\tau(x+2[\alpha_2]_{\rm o}+2[\alpha_3]_{\rm o})}{\tau(x)}\right.
\nonumber
\\
\hphantom{\frac{\tau\left(x+2\sum\limits_{i=1}^{4}[\alpha_i]_{\rm o}\right)}{\tau(x)}=}{}
-\frac{\alpha_{24}}{\tilde{\alpha}_{24}}\frac{\tau(x+2[\alpha_2]_{\rm o}+2[\alpha_4]_{\rm o})}{\tau(x)}
\frac{\alpha_{13}}{\tilde{\alpha}_{13}}\frac{\tau(x+2[\alpha_1]_{\rm o}+2[\alpha_3]_{\rm o})}{\tau(x)}\nonumber
\\
\left.
\hphantom{\frac{\tau\left(x+2\sum\limits_{i=1}^{4}[\alpha_i]_{\rm o}\right)}{\tau(x)}=}{}
+\frac{\alpha_{34}}{\tilde{\alpha}_{34}}\frac{\tau(x+2[\alpha_3]_{\rm o}+2[\alpha_4]_{\rm o})}{\tau(x)}\frac{\alpha_{12}
}{\tilde{\alpha}_{12}}\frac{\tau(x+2[\alpha_1]_{\rm o}+2[\alpha_2]_{\rm o})}{\tau(x)}\right\}.
\label{eq:56}
\end{gather}
\end{example}

 As is proved in Proposition~\ref{proposition8}, equation~\eqref{eq:56} can be derived from equation~\eqref{eq:55}.
\begin{theorem}\label{theorem3}
The four term equation~\eqref{eq:55} is equivalent to the BKP hierarchy~\eqref{eq:50}.
\end{theorem}

We prove this theorem in a~similar way to the case of the KP hierarchy.

In order to prove the theorem, we use Pfaf\/f\/ians.
The def\/inition and notation of Pfaf\/f\/ians are reviewed in Appendix~\ref{appendixA}.

Let us def\/ine the components of Pfaf\/f\/ians by
\begin{gather*}
(0,j)=\frac{\tau(x+2[\alpha_j]_{\rm o})}{\tau(x)},\qquad \hskip1cm(i,j)=\frac{\alpha_{ij}}{\tilde{\alpha}_{ij}}
\frac{\tau(x+2[\alpha_i]_{\rm o}+2[\alpha_j]_{\rm o})}{\tau(x)}.
\end{gather*}
Then it is possible to rewrite~\eqref{eq:55} and~\eqref{eq:56} as
\begin{gather}
\frac{\tau\left(x+2\sum\limits_{i=1}^{3}[\alpha_i]_{\rm o}\right)}{\tau(x)}=A_{123}(0,1,2,3),
\label{eq:58}
\\
\frac{\tau\left(x+2\sum\limits_{i=1}^{4}[\alpha_i]_{\rm o}\right)}{\tau(x)}=A_{1234}(1,2,3,4),
\label{eq:59}
\end{gather}
respectively.

The following proposition can be proved in a~similar manner to Proposition~\ref{proposition1}.
\begin{proposition}\label{proposition7}
The BKP hierarchy~\eqref{eq:50} is equivalent to~\eqref{eq:53} and~\eqref{eq:54}.
\end{proposition}
\begin{proposition}\label{proposition8}
The following equations are implied by~\eqref{eq:55}:
\begin{gather}
\frac{\tau\left(x+2\sum\limits_{i=1}^{n}[\alpha_i]_{\rm o}\right)}{\tau(x)}=A_{1\dots n}(0,1,2,\dots,n),\qquad n\ge3, \ \text{odd},
\label{eq:66}
\\
\frac{\tau\left(x+2\sum\limits_{i=1}^{n}[\alpha_i]_{\rm o}\right)}{\tau(x)}=A_{1\dots n}(1,2,\dots,n),\qquad n\ge4, \ \text{even}.
\label{eq:67}
\end{gather}
\end{proposition}

\begin{proof}
First we prove that~\eqref{eq:55} implies~\eqref{eq:56}.
Shift $x$ in~\eqref{eq:55} as $x\rightarrow x+2[\alpha_4]_{\rm o}$:
\begin{gather}
\frac{\tau\left(x+2\sum\limits_{i=1}^{4}[\alpha_i]_{\rm o}\right)}{\tau(x+2[\alpha_4]_{\rm o})}\nonumber\\
\qquad{}
=A_{123}\frac{\tau(x)^2}{\tau(x+2[\alpha_4]_{\rm o})^2}\left\{\frac{\tau(x+2[\alpha_1]_{\rm o}+2[\alpha_4]_{\rm o})}{\tau(x)}
\frac{\alpha_{23}}{\tilde{\alpha}_{23}}\frac{\tau(x+2[\alpha_2]_{\rm o}+2[\alpha_3]_{\rm o}+2[\alpha_4]_{\rm o})}{\tau(x)}
\right.
\nonumber
\\
\qquad\quad{} -\frac{\tau(x+2[\alpha_2]_{\rm o}+2[\alpha_4]_{\rm o})}{\tau(x)}\frac{\alpha_{13}}{\tilde{\alpha}_{13}}
\frac{\tau(x+2[\alpha_1]_{\rm o}+2[\alpha_3]_{\rm o}+2[\alpha_4]_{\rm o})}{\tau(x)}\nonumber
\\
 \left.
 \qquad\quad{} +\frac{\tau(x+2[\alpha_3]_{\rm o}+2[\alpha_4]_{\rm o})}{\tau(x)}\frac{\alpha_{12}}{\tilde{\alpha}_{12}}
\frac{\tau(x+2[\alpha_1]_{\rm o}+2[\alpha_2]_{\rm o}+2[\alpha_4]_{\rm o})}{\tau(x)}\right\}.
\label{eq:68}
\end{gather}
Use~\eqref{eq:58} to rewrite $\tau(x+2[\alpha_{i_1}]_{\rm o}+2[\alpha_{i_2}]_{\rm o}+2[\alpha_{i_3}]_{\rm o})$
in~\eqref{eq:68}.
Then we get~\eqref{eq:59}.
We prove~\eqref{eq:66} by induction on $n$.
The case $n=3$ is obvious.
Suppose that~\eqref{eq:66} holds in case of $n$:
\begin{gather}
\frac{\tau\left(x+2\sum\limits_{i=1}^{n}[\alpha_i]_{\rm o}\right)}{\tau(x)}=A_{1\dots n}(0,1,2,\dots,n)=A_{1\dots n}{\rm Pf}\, {\mathcal A},
\label{eq:69}
\end{gather}
where ${\mathcal A}=(a_{ij})_{0\le i,j\le n}$ is a~skew-symmetric matrix,
\begin{gather*}
a_{ij}=
\begin{cases}\dfrac{\tau(x+2[\alpha_j]_{\rm o})}{\tau(x)}, &  i=0,
\vspace{2mm}\\
\dfrac{\alpha_{ij}}{{\tilde{\alpha}}_{ij}}\dfrac{\tau(x+2[\alpha_i]_{\rm o}+2[\alpha_j]_{\rm o})}{\tau(x)}
,&  i\neq0,\quad  i<j.
\end{cases}
\end{gather*}
In~\eqref{eq:69} we shift $x$ as
\begin{gather*}
x\rightarrow x+2[\alpha_{n+1}]_{\rm o}+2[\alpha_{n+2}]_{\rm o}.
\end{gather*}
Then we have, using~\eqref{eq:58} and~\eqref{eq:59},
\begin{gather}
\frac{\tau\left(x+2\sum\limits_{i=1}^{n+2}[\alpha_i]_{\rm o}\right)}{\tau(x)}=A_{1\dots n+2}(n+1,n+2)^{-\frac{n-1}{2}}{\rm Pf}\,
{\mathcal B},
\label{eq:71}
\end{gather}
where ${\mathcal B}=(b_{ij})_{0\le i<j\le n}$ is a~skew-symmetric matrix given by
\begin{gather*}
b_{ij}=(n+1,n+2,i,j),\qquad  i<j .
\end{gather*}

By the analogue of the Sylvester' theorem for Pfaf\/f\/ians (Appendix~\ref{appendixB}), we have
\begin{gather*}
{\rm Pf}((1,2,\dots,2r,i,j))_{2r+1\le i<j\le2m}=(1,2,\dots,2r)^{m-r-1}(1,2,\dots,2m).
\end{gather*}
Consider the case $r=1$ and $2m=n+3$ of this formula:
\begin{gather}
{\rm Pf}((n+1,n+2,i,j))_{0\le i<j\le n}=(n+1,n+2)^{\frac{n-1}{2}}(n+1,n+2,0,\dots,n).
\label{eq:80}
\end{gather}
Substituting~\eqref{eq:80} to~\eqref{eq:71}, we get
\begin{gather*}
\frac{\tau\left(x+2\sum\limits_{i=1}^{n+2}[\alpha_i]_{\rm o}\right)}{\tau(x)}=A_{1\dots n+2}(0,1,\dots,n+2).
\end{gather*}
The case of even $n$ is similarly proved.
\end{proof}

\begin{proposition}\label{proposition9}
The Pl\"{u}cker's relations for Pfaffians of the right hand side of~\eqref{eq:66} and~\eqref{eq:67}
give the addition formulae~\eqref{eq:53} and~\eqref{eq:54} respectively.
\end{proposition}

\begin{proof}
Using the Pl\"{u}cker's relations~\eqref{eq:500} and~\eqref{eq:501} for Pfaf\/f\/ians given in Appendix~\ref{appendixC},
the proposition can easily be checked by direct calculations.
\end{proof}

\appendix

\section{Pfaf\/f\/ians} \label{appendixA}

Let ${\mathcal A} =(a_{ij})_{1\le i,j\le2m}$ be a~skew-symmetric matrix.
Then the Pfaf\/f\/ian ${\rm Pf}\, {\mathcal A}$ is def\/ined by
\begin{gather*}
\det{\mathcal A}=({\rm Pf}\,{\mathcal A})^2,\qquad {\rm Pf}\,{\mathcal A}=a_{12}a_{34}\cdots a_{2m-1,2m}+\cdots.
\end{gather*}
Following~\cite{H1} we denote ${\rm Pf}\, {\mathcal A}$ by $(1,2,3,\dots,2m)$:
\begin{gather*}
{\rm Pf}\,{\mathcal A}=(1,2,3,\dots,2m).
\end{gather*}
It is directly def\/ined by
\begin{gather*}
(1,2,3,\dots,2m)=\sum{\rm sgn}(i_1,\dots,i_{2m})\cdot(i_1,i_2)(i_3,i_4)\cdots(i_{2m-1},i_{2m}
),\qquad (i,j)=a_{ij},
\end{gather*}
where the sum is over all permutations of $(1,\dots,2m)$ such that
\begin{gather*}
i_1<i_3<\cdots<i_{2m-1},\qquad i_1<i_2,\dots,i_{2m-1}<i_{2m},
\end{gather*}
and ${\rm sgn}(i_1,\dots,i_{2m})$ is the signature of the permutation $(i_1,\dots,i_{2m})$.
The Pfaf\/f\/ian can be expanded as
\begin{gather*}
(1,2,3,\dots,2m)=\sum_{j=2}^{2m}(-1)^j(1,j)(2,3,\dots,\hat{j},\dots,2m).
\end{gather*}
For example, in the case of $m=2$,
\begin{gather*}
(1,2,3,4)=(1,2)(3,4)-(1,3)(2,4)+(1,4)(2,3).
\end{gather*}

\section{Sylvester's theorem for determinants and Pfaf\/f\/ians} \label{appendixB}

The following theorem is known as
Sylvester's theorem.
\begin{theorem}\label{theorem4}
Let $r\le m$, ${\mathcal A}=(a_{ij})_{1\le i,j\le m}$ and ${\mathcal A}_r=(a_{ij})_{1\le i,j\le r}$.
Set
\begin{gather*}
{\mathcal B}=(b_{ij})_{r+1\le i,j\le m},
\qquad
b_{ij}=\det\left(
\begin{matrix}a_{11}&\hdots&a_{1r}&a_{1j}
\\
\vdots&\ddots&\vdots&\vdots
\\
a_{r1}&\hdots&a_{rr}&a_{rj}
\\
a_{i1}&\hdots&a_{ir}&a_{ij}
\end{matrix}
\right).
\end{gather*}
Then we get
\begin{gather*}
\det{\mathcal B}=(\det{\mathcal A}_r)^{m-r-1}\det{\mathcal A}.
\end{gather*}
\end{theorem}

  Let ${\mathcal A}=(a_{ij})_{1\le i,j\le2m}$ be a~skew-symmetric matrix and set $(i,j)=a_{ij}$.
For $r\le m$ let ${\mathcal P} =(p_{ij})_{2r+1\le i,j\le2m}$, $p_{ij}=(1,2,\dots,2r,i,j)$ and $I_r=\{1,2,\dots,2r\}$.
In general, for a~subset $I\subset\{1,2,\dots,2m\}$ we set ${\mathcal A}(I)=(a_{ij})_{i,j\in I}$ and for $i<j$, $k<l$
we denote by ${\mathcal A}_{kl}^{ij}$ be the square matrix of degree $2(m-1)$ which is obtained from ${\mathcal A}$ by
removing $i$-th and $j$-th rows, $k$-th and $l$-th columns.

\begin{theorem}[\cite{H2}]\label{theorem5} For $r\le m$ the following identity holds:
\begin{gather*}
{\rm Pf}\,{\mathcal P}=({\rm Pf}\,{\mathcal A}(I_r))^{m-r-1}{\rm Pf}\,{\mathcal A}.
\end{gather*}
\end{theorem}

\section{The Pl\"{u}cker relation for Pfaf\/f\/ians}\label{appendixC}

 There exist analogues of the Pl\"{u}cker's
relations for Pfaf\/f\/ians~\cite{O1}.
They are given by
\begin{gather}
\sum_{l=1}^L(-1)^l(i_1,\dots,i_K,j_l)(j_1,\dots,\hat{j}_l,\dots,j_L)
\nonumber
\\
\qquad
{}+\sum_{k=1}
^K(-1)^k(i_1,\dots,\hat{i}_k,\dots,i_K)(j_1,\dots,j_L,i_k)=0,
\label{eq:75}
\end{gather}
where $K$ and $L$ are odd.
We understand that $(\varnothing)=1$.

For $n$ odd, taking $K=1$, $L=n$, $i_1=0$ and $j_1,\dots,j_n\neq0$ in~\eqref{eq:75}, we get
\begin{gather}
\sum_{l=1}^n(-1)^{l-1}(0,j_l)(j_1,\dots,\hat{j}_l,\dots,j_n)-(0,j_1,\dots,j_n)=0.
\label{eq:500}
\end{gather}
For $n$ even, setting $K=1$, $L=n-1$ and $i_1\neq0$ in~\eqref{eq:75}, we have
\begin{gather}
\sum_{l=1}^{n-1}(-1)^{l-1}(i_1,j_l)(j_1,\dots,\hat{j}_l,\dots,j_{n-1})-(i_1,j_1,\dots,j_{n-1})=0.
\label{eq:501}
\end{gather}

\subsection*{Acknowledgements} I would like to thank Masatoshi Noumi and Takashi Takebe for permitting me
to see the manuscript~\cite{NT1} prior to its publication.
I also thank Kanehisa Takasaki and Takashi Takebe for insightful comments and their interests in the
present work.
I also thank Yasuhiro Ohta and Soichi Okada for important comments about Pfaf\/f\/ians and informing me
about reference~\cite{H2}.
Finally I am deeply grateful to Atsushi Nakayashiki for much advice.
This research is supported by JSPS Grant-in-Aid for Scientif\/ic Research~(B) No.~23340037.

\pdfbookmark[1]{References}{ref}
\LastPageEnding


\begin{thebibliography}{99}
\footnotesize\itemsep=0pt

\bibitem{BS1}
Bobenko A.I., Suris Yu.B., Discrete dif\/ferential geometry. Integrable structure,
  \href{http://dx.doi.org/10.1007/978-3-7643-8621-4}{\textit{Graduate Studies in Mathematics}}, Vol.~98, Amer. Math. Soc.,
  Providence, RI, 2008.

\bibitem{BEL1}
Buchstaber V.M., Enolski V.Z., Leykin D.V., Kleinian functions, hyperelliptic
  Jacobians and applications, \textit{Rev. Math and Math. Phys.} \textbf{10}
  (1997), no.~2, 1--125, \href{http://arxiv.org/abs/solv-int/9603005}{solv-int/9603005}.

\bibitem{DJM1-I}
Date E., Jimbo M., Miwa T., Method for generating discrete soliton
  equations.~{I}, \href{http://dx.doi.org/10.1143/JPSJ.51.4116}{\textit{J.~Phys. Soc. Japan}} \textbf{51} (1982), 4116--4124.

\bibitem{DJM1-II}
Date E., Jimbo M., Miwa T., Method for generating discrete soliton
  equations.~{II}, \href{http://dx.doi.org/10.1143/JPSJ.51.4125}{\textit{J.~Phys. Soc. Japan}} \textbf{51} (1982), 4125--4131.

\bibitem{DJM1-V}
Date E., Jimbo M., Miwa T., Method for generating discrete soliton
  equations.~{V}, \href{http://dx.doi.org/10.1143/JPSJ.52.766}{\textit{J.~Phys. Soc. Japan}} \textbf{52} (1982), 766--771.

\bibitem{DJKM1}
Date E., Kashiwara M., Jimbo M., Miwa T., Transformation groups for soliton
  equations, in Nonlinear In\-tegrable Systems~-- Classical Theory and Quantum
  Theory ({K}yoto, 1981), World Sci. Publishing, Singapore, 1983, 39--119.

\bibitem{EEG1}
Eilbeck J.C., Enolski V.Z., Gibbons J., Sigma, tau and {A}belian functions of
  algebraic curves, \href{http://dx.doi.org/10.1088/1751-8113/43/45/455216}{\textit{J.~Phys.~A: Math. Theor.}} \textbf{43} (2010),
  455216, 20~pages, \href{http://arxiv.org/abs/1006.5219}{arXiv:1006.5219}.

\bibitem{F1}
Fay J.D., Theta functions on {R}iemann surfaces, \textit{Lecture Notes in
  Mathematics}, Vol.~352, Springer-Verlag, Berlin, 1973.

\bibitem{H2}
Hirota R., Generalizations of determinant identities by Pfaf\/f\/ian, in
  Mathematical Theories and Applications of Nonlinear Waves and Nonlinear
  Dynamics, Research Institute for Applied Mechanics, Kyushu University, 2004, 148--156.

\bibitem{H1}
Hirota R., The direct method in soliton theory, \href{http://dx.doi.org/10.1017/CBO9780511543043}{\textit{Cambridge Tracts in
  Mathematics}}, Vol.~155, Cambridge University Press, Cambridge, 2004.


\bibitem{IKMO1}
Inoguchi J., Kajiwara K., Matsuura N., Ohta Y., Explicit solutions to the
  semi-discrete modif\/ied KdV equation and motion of discrete plane curves,
  \href{http://arxiv.org/abs/1108.1328}{arXiv:1108.1328}.

\bibitem{JM1}
Jimbo M., Miwa T., Solitons and inf\/inite-dimensional {L}ie algebras,
  \href{http://dx.doi.org/10.2977/prims/1195182017}{\textit{Publ. Res. Inst. Math. Sci.}} \textbf{19} (1983), 943--1001.

\bibitem{MJD1}
Miwa T., Jimbo M., Date E.,
Solitons. Dif\/ferential equations, symmetries and inf\/inite-dimensional algebras,
\textit{Cambridge Tracts in Mathematics}, Vol.~135, Cambridge University Press, Cambridge, 2000.

\bibitem{Mac}
Macdonald I.G., Symmetric functions and {H}all polynomials, 2nd ed., \textit{Oxford
  Mathematical Monographs}, The Clarendon Press, Oxford University Press, New
  York, 1995.

\bibitem{M1}
Miwa T., On {H}irota's dif\/ference equations, \href{http://dx.doi.org/10.3792/pjaa.58.9}{\textit{Proc. Japan Acad. Ser.~A
  Math. Sci.}} \textbf{58} (1982), 9--12.

\bibitem{N1}
Nakayashiki A., Sigma function as a tau function, \href{http://dx.doi.org/10.1093/imrn/rnp135}{\textit{Int. Math. Res. Not.}}
  {\bf 2010} (2010), no.~3, 373--394, \href{http://arxiv.org/abs/0904.0846}{arXiv:0904.0846}.

\bibitem{NT1}
Noumi M., Takebe T., Algebraic analysis of integrable hierarchies, {i}n
  preparation.

\bibitem{O1}
Ohta Y., Bilinear theory of solitons with {P}faf\/f\/ian labels,
  \textit{S\=urikaisekikenky\=usho K\=oky\=uroku}  (1993), no.~822, 197--205.

\bibitem{R1}
Raina A.K., Fay's trisecant identity and conformal f\/ield theory, \href{http://dx.doi.org/10.1007/BF01256498}{\textit{Comm.
  Math. Phys.}} \textbf{122} (1989), 625--641.

\bibitem{SS1}
Sato M., Sato Y., Soliton equations as dynamical systems on inf\/inite
  dimensional Grassmann manifold, in Nonlinear PDE in Applied Science,
  \textit{North-Holland Math. Stud.}, Vol.~81, Editors H.~Fujita, P.~Lax,
  G.~Strang, Tokyo, 1982, 259--271.

\bibitem{T1}
Takasaki K., Dif\/ferential {F}ay identities and auxiliary linear problem of
  integrable hierarchies, in Exploring New Structures and Natural Constructions
  in Mathematical Physics, \textit{Adv. Stud. Pure Math.}, Vol.~61, Math. Soc.
  Japan, Tokyo, 2011, 387--441, \href{http://arxiv.org/abs/0710.5356}{arXiv:0710.5356}.

\bibitem{T2}
Takasaki K., Dispersionless {H}irota equations of two-component {BKP}
  hierarchy, \href{http://dx.doi.org/10.3842/SIGMA.2006.057}{\textit{SIGMA}} \textbf{2} (2006), 057, 22~pages,
  \href{http://arxiv.org/abs/nlin.SI/0604003}{nlin.SI/0604003}.


\bibitem{TT1}
Takasaki K., Takebe T., Integrable hierarchies and dispersionless limit,
  \href{http://dx.doi.org/10.1142/S0129055X9500030X}{\textit{Rev. Math. Phys.}} \textbf{7} (1995), 743--808,
  \href{http://arxiv.org/abs/hep-th/9405096}{hep-th/9405096}.

\bibitem{Te1}
Teo L.P., Fay-like identities of the {T}oda lattice hierarchy and its
  dispersionless limit, \href{http://dx.doi.org/10.1142/S0129055X06002838}{\textit{Rev. Math. Phys.}} \textbf{18} (2006),
  1055--1073, \href{http://arxiv.org/abs/nlin.SI/0606059}{nlin.SI/0606059}.

\end{thebibliography}
\end{document}